\documentclass[conference]{IEEEtran}
\IEEEoverridecommandlockouts
\usepackage{cite}
\usepackage{amsmath,amssymb,amsfonts}
\usepackage{algorithmic}
\usepackage{graphicx}
\usepackage{multirow}
\usepackage{makecell}
\usepackage{textcomp}
\usepackage{xcolor}
\usepackage[ruled,linesnumbered]{algorithm2e}
\def\BibTeX{{\rm B\kern-.05em{\sc i\kern-.025em b}\kern-.08em
    T\kern-.1667em\lower.7ex\hbox{E}\kern-.125emX}}
\begin{document}

\title{Code Clone Detection based on Event Embedding and Event Dependency\\
}

\author{\IEEEauthorblockN{Cheng Huang, Hui Zhou*, Chunyang Ye*\thanks{*Corresponding author}, Bingzhuo Li}
	\IEEEauthorblockA{\textit{School of Computer Science and Technology} \\
		\textit{Hainan University}\\
		Haikou, China \\
		\{chenghuang, bzli, huizhou, cyye\}@hainanu.edu.cn}
}

\maketitle

\begin{abstract}
	
The code clone detection method based on semantic similarity has important
	value in software engineering tasks (e.g., software evolution, software
	reuse). Traditional  code clone detection technologies pay more
	attention to the similarity of code at the syntax level, and less attention
	to the semantic similarity of the code. As a result, candidate codes
	similar in semantics are ignored. To address this issue, we propose a
	code clone detection method based on semantic similarity. By treating
	code as a series of interdependent events that occur continuously, we
	design a model namely EDAM to encode code semantic information based on
	event embedding and event dependency.  The EDAM model uses the event
	embedding method to model the execution characteristics of program
	statements and the data dependence information between all statements.
	In this way, we can embed the program semantic information into a
	vector and use the vector to detect codes similar in semantics.
	Experimental results show that the performance of our EDAM model is
	superior to state-of-the-art open source models for code clone
	detection.

\end{abstract}


\section{Introduction}

Code clone detection technology is important for many software engineering
tasks (e.g., software evolution, software reuse). Existing code clone detection methods are mainly divided into three categories, namely text-based, syntax-based and
semantic-based clone detection methods~\cite{Mu2019A}. These code clone dection
methods play an important role in program understanding, plagiarism detection,
copyright protection, code compression, software evolution analysis, code
quality analysis, bug detection, and anti-virus. The core functionality of the
code detection model is to calculate the similarity of the code. Given a
certain target code fragment, the system will first calculates the similarity
between the target code fragment and all the code fragments in the database,
and then returns the result according to the similarity between the codes.
The similarity between codes can be classified into four
levels~\cite{Roy2007A}. Type-1 similarity means that the two pieces of code are
identical except for the differences in spaces, layout and comments.  Type-2
similarity means the code pairs are identical except for the variable name,
type name, and function name. Type-3 similarity means that there are several
additions and deletions of statements, and the use of different identifiers,
text, types, spaces, layout and comments in the code pairs, but they are still
similar.  Type-4 similarity detects code pairs that are functionally similar,
but they are different in text or syntax.
Type-3 and Type-4 similarity can also be further divided into
three categories based on their syntactical similarity values~\cite{Sva2014To}:
Strongly Type-3, similarity in range between 0.7 and 1.0, Moderately Type-3,
similarity in range between 0.5 and 0.7, and Weakly Type-3 which similarity in
range between 0 and 0.5. Weakly Type-3 clone codes are also regarded as a Type-4 clone codes.

In practice, it is easy to detect Type-1 clone samples, but it is the most
difficult to detect clone samples of Type-4. The model proposed in
this paper mainly focuses on the detection of Type-3/4 clone samples. The key
idea of code clone detection methods is to extract some information from code
fragment, and then use this type of information to identify the semantically
similar code fragments. According to the type of information used, these
approaches can be classified into Text-based approaches, Token-based
approaches, Tree-based approaches, Metric-based approaches, and Graph-based
approaches. Existing methods mainly focus on the static characteristics of the
source code, and seldom consider the dynamic characteristics of the code.

\begin{figure*}[hbt]
	\vspace{-3mm}
	\centering
	\includegraphics[width=\linewidth]{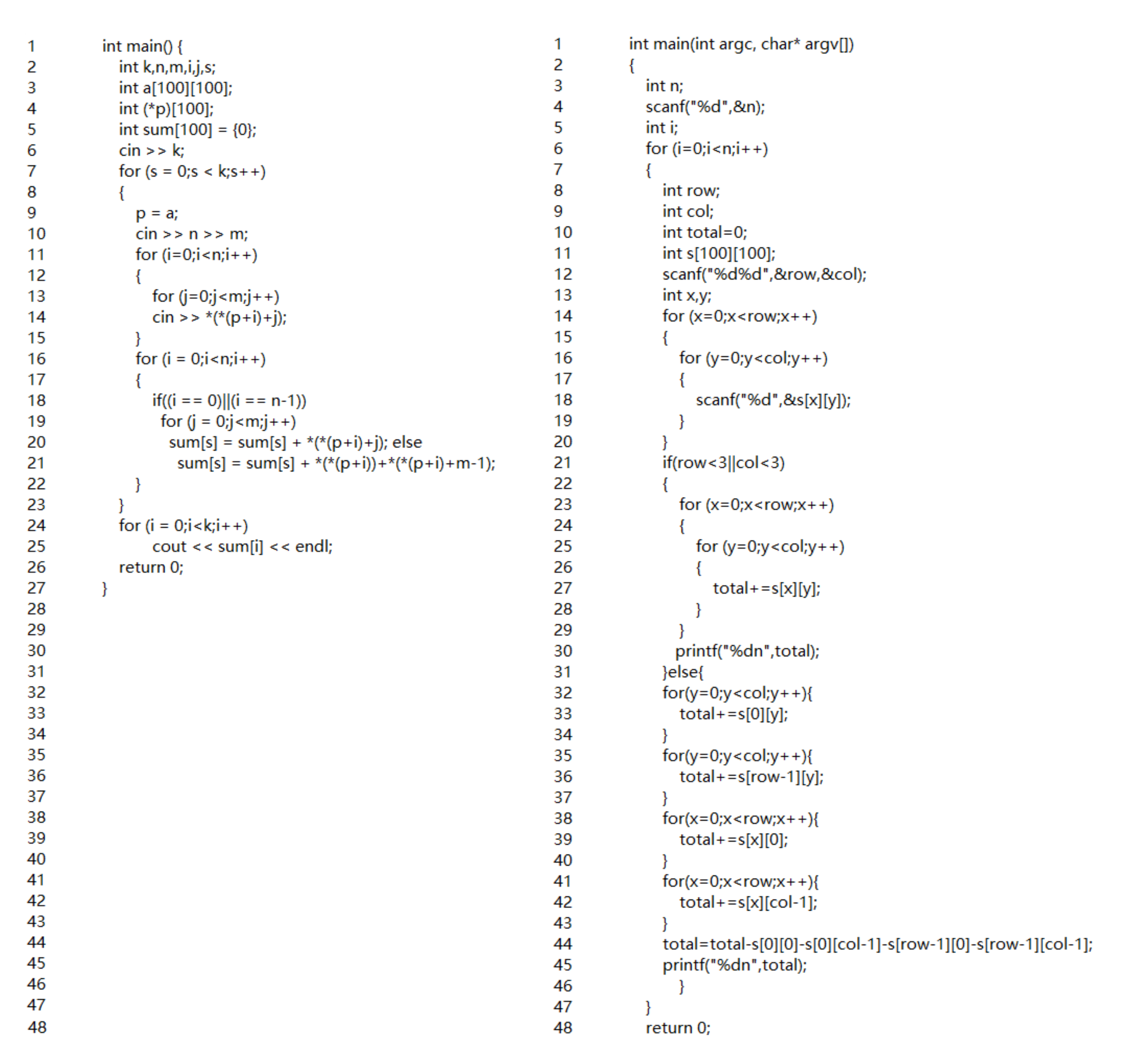}
	\caption{A clone code pair}
	\label{fig:code_compare}
	\vspace{-3mm}
\end{figure*}
As shown in Fig.~\ref{fig:code_compare}, this code pair is selected from the data set used in this paper. They are highly similar in function, but low in grammatical similarity.
The traditional code clone detection method cannot detect the clone sample shown in Fig.~\ref{fig:code_compare}.By contrast, the CSEM model we proposed before uses event embedding to model the semantic information of a single statement, and then uses the GAT network to model the control flow information of the program. The reason for modeling control flow is that we believe that control flow of code reflects the semantic dependencies between code statements. Through the above steps, the CSEM model can encode code semantic information.
The CSEM model uses the GAT network to model the dependencies of code statements. This approach may have the following problems. First of all, the code statements that are adjacent to each other in the control flow graph may not have interdependence. Since the GAT network can only model the adjacent nodes in the control flow graph, the wrong dependency is captured. For example, as shown in left part of Fig.~\ref{fig:code_compare}, there is no dependency between the statements 5 and 6, but they are adjacent in the control flow graph. Second, limited by the structure and training cost of the GAT network, we can only model the statement dependency on the first-order adjacent nodes in the control flow graph.
Therefore, it is difficult to capture the relationships between program statements that have semantic dependencies, but are far apart in the source code (usually not adjacent in the control flow graph). For example, as shown in right part of Fig.~\ref{fig:code_compare}, statement 11 and statement 27 have a dependency relationship of variable s, but the GAT network usually cannot model this type of dependency relationship.
In order to solve the above-mentioned limitations of the CSEM model, we propose an EDAM model based on event-dependent graphs. Compared with the CSEM model, the EDAM model directly focuses on the semantic dependencies between statements.Therefore, the EDAM model solves the limitations of the CSEM model we mentioned above. In our experiments, the EDAM model also has better prediction accuracy.

The advantage of our method is that it can model the dynamic semantics of  the
code. The key-idea of this paper is that a program can be regarded as a series
of interdependent events that occur continuously. Therefore, by modeling these
events, we can extract the dynamic semantic information of the program. The
challenges of this method are how to extract the dynamic semantic information
of the code from the source code file, and how to enable the model to use this
information to calculate the semantic similarity of two pieces of code. In
order to address the above challenges, we propose a model EDAM that uses event
embedding to model the execution characteristics of the program's statements
and the data dependence (event dependency) between different statements. We
perform event dependency analysis on the source code and transform it into an
event dependency graph. The event dependency graph describes the statement
execution characteristics of the code and the event dependency relationship
between different statements. After that, we input the event dependency graph
into the event dependency execution engine, which will analyzes and calculate
the event dependency graph, and finally output the vector representation of
the program. This vector contains the event-dependent semantic information of
the program, so it can be used for semantic-based code clone detection.

The main contributions of this paper are as follows:
\begin{itemize}
	
	\item We propose a code clone detection method, which regards the
		program as a series of continuous interdependent events. We use
		this key-idea to model the dynamic semantic information of the
		code. Then we use the embedded vector to measure the semantic
		similarity of the code. Experimental results show that our
		method has certain advantages over the comparison models in
		Type-3/4 code clone detection.

	\item This article develops a tool chain for the model. The tools in
		the tool chain can generate the event dependency graph of the
		program through the event dependency analyzer. Then event
		embedding execution engine will calculates code semantic vector
		by using event dependency graph. The code semantic vector can
		be used to perform code clone detection tasks. 

\end{itemize}

The rest of this paper is organized as follows: Section.~\ref{section:related} reviews the
works in code clone detection. Section.~\ref{section:preliminary} introduces
some preliminary about clone detection. Section.~\ref{section:model} describes the
EDAM model in detail. Section.~\ref{section:evaluation} evaluates our EDAM
model through experiments. Section.~\ref{section:conclusion} summarizes the
work of this paper and highlights some future work directions.

\section{related work}\label{section:related}

The code clone detection model has important value in software engineering tasks such
as bug detection, copyright protection, software evolution analysis, and
anti-virus. Therefore, the study of code clone detection has attracted more
attention in recent years. For example, Kim et al. present a model called VUDDY
to find out the code pairs that contain the similar risk of
error~\cite{Kim2018Soft,Kim2017VUDDY}. The core function of the code clone
detection model is to convert the code into its vector representation, and then
calculate the similarity based on the code embedding vector, thereby selecting
the most suitable one or more pieces of code from the candidate code segments
as the result. According to the different types of code information used in the
code embedding model, existing methods can be divided into the following
categories: text-based methods~\cite{Du2000Lan, Jo1993Identify,Jo1994Substring,
Ra2017Using}, token-based methods~\cite {Ni2017Sca, Ba1992Pro, Ka2002CC},
metric based methods~\cite {Tsu2016Asses, Sva2017Fast}, tree-based
methods~\cite{Ko2006Clone, Cho2015Hask}, and graph-based methods~\cite
{Krinke2001Identifying, Wang2017CCSharp,li2020semantic}.

According to the difficulty of code clone detection, it can be divided into
Type-1, Type-2, Type-3 and Type-4 code clone detection~\cite{Roy2007A}. The
Type-1/2 code clone detection is simple, while the Type-3/4 code clone
detection is more difficult. SourcerCC \cite{Sajn2016Sour}, CloneWork
\cite{Sva2017Fast}, Nicad \cite{Roy2008Nicad} and CCLearner \cite{Li2017CC}
are the most popular models in Type-3 code clone detection. SourcerCC and
CloneWork are both hybrid models based on token and index. Nicad is a model
that embeds programs based on textual information. Nicard filters and
normalizes the code fragments to eliminate the interference of irrelevant
factors on the prediction results of the model. CCLearner works on the lexical
level, divides program tokens into eight categories and then represents
programs as token-frequency list vectors. Then, for each code pair, CCLearner
computes a similarity score between the token-frequency lists to identify the
similar code pairs. Deckard utlizes abstract syntax tree to embed the program,
and then it clusters the embedding vectors of the program to identify similar
code segments.

For type-4 detection, Gabel et al. detect semantic similarity by augmenting
Deckard \cite{Jiang2007Deckard} with a step to generate vectors for semantic
similarity codes \cite{Gabel2008Sca}. Jiang et al. propose a method to detect
semantic similarity codes  by executing code fragments against random inputs
\cite{jiang2009Auto}. Wei and Li use LSTM to generate the vector expression of
the code segment, and then calculate the hamming distance of them with the hash
function to determine whether they are similiar pairs \cite{Hu2017Su}. The
pairs with the smaller hamming distance are classified as similiar pairs.
DeepSim encodes the control flow and data flow of a code fragment into a
semantic matrix, based on which a deep neural network is designed to measure
code functional similarity \cite{Zhao2018Deep}. Oreo represents the code
characteristics using 24 metrics, including the number of variables declared
\cite{Sai2018Oreo}. Then, they train a binary classifier using a vector of 48
dimensions, which corresponds to a pair of code fragments as an input into a
symmetrically structured Siamese network. Li et al. propose the CSEM model
\cite{li2020semantic}, which uses the GAT network to model the program control
flow information, so that the semantics of the program can be embedded.

Different from SourcerCC, Nicad and CloneWork, EDAM is a code clone detection
model based on the event dependency graph (a data structure proposed in this
paper, C.f. Section.~\ref{section:model}), which has a better ability to model
code semantics. Also different from the CCLearner and Oreo models, our EDAM
model does not need to manually select the mertrics that need to be collected
from the code. Different from the DeepSim model, the EDAM model  uses an event
dependency graph instead of a control flow graph to capture the execution
semantics of the code. Different from the CSEM model, the EDAM
model introduces the concept of event dependency, which can model the data flow
information of the program. Compared with the CDLH model that uses the LSTM
network to model the program abstract syntax tree, the EventTransformer used in
the EDAM model has a stronger ability to model program semantics. In summary,
the EDAM model proposed in this paper can more effectively model the semantic
information of the code.

\section{preliminary}\label{section:preliminary}
\subsection{event embedding}
\begin{figure}[hbt]
	\vspace{-3mm}
	\centering
	\includegraphics[width=\linewidth]{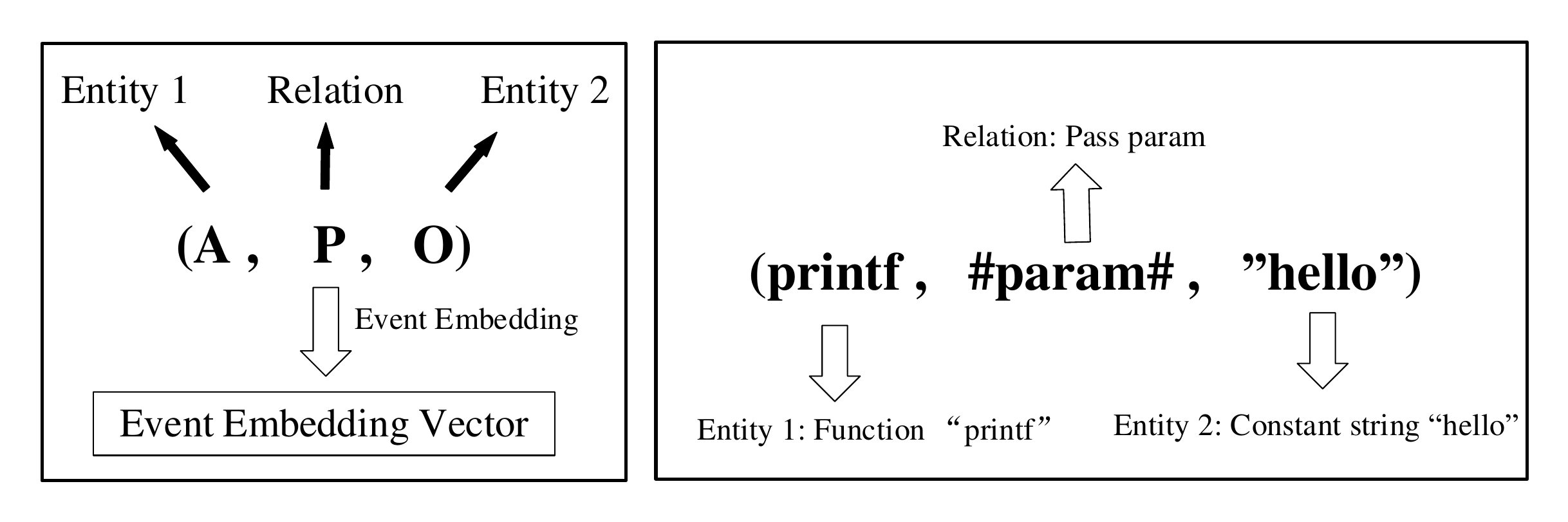}
	\caption{Event embedding in program}
	\label{fig:event_embedding_program}
	\vspace{-3mm}
\end{figure}

We describe the definition of event embedding as follows: Consider a triple
$(A, P, O)$ as shown in Fig. \ref{fig:event_embedding_program}, where $A$ is
entity 1, $O$ is entity 2, and $P$ is the relationship between two entities.
The purpose of event embedding is to convert triples $(A, P, O)$ into a vector
$X$.

The program can be regarded as a series of interdependent events that occur
continuously, so we can use the event embedding method to model the semantics
of the program. As shown in  Fig. \ref{fig:event_embedding_program}, the
function statement printf("hello") can be regarded as an event, in which the
function printf is regarded as entity 1, the string constant "hello" as the
function parameter is regarded as entity 2, and the relationship "passing
parameters" between the two entities is regarded as the relationship $P$. Therefore, the
event print ("hello") can be converted into a vector via event embedding. It
should be noted that in  Fig. \ref{fig:event_embedding_program}, “\#param\#”
wrapped by “\#” represents the relationship between entities.

\subsection{event dependency}

\begin{figure}[hbt]                                 
	\centering                                 
	\includegraphics[width=0.7\linewidth]{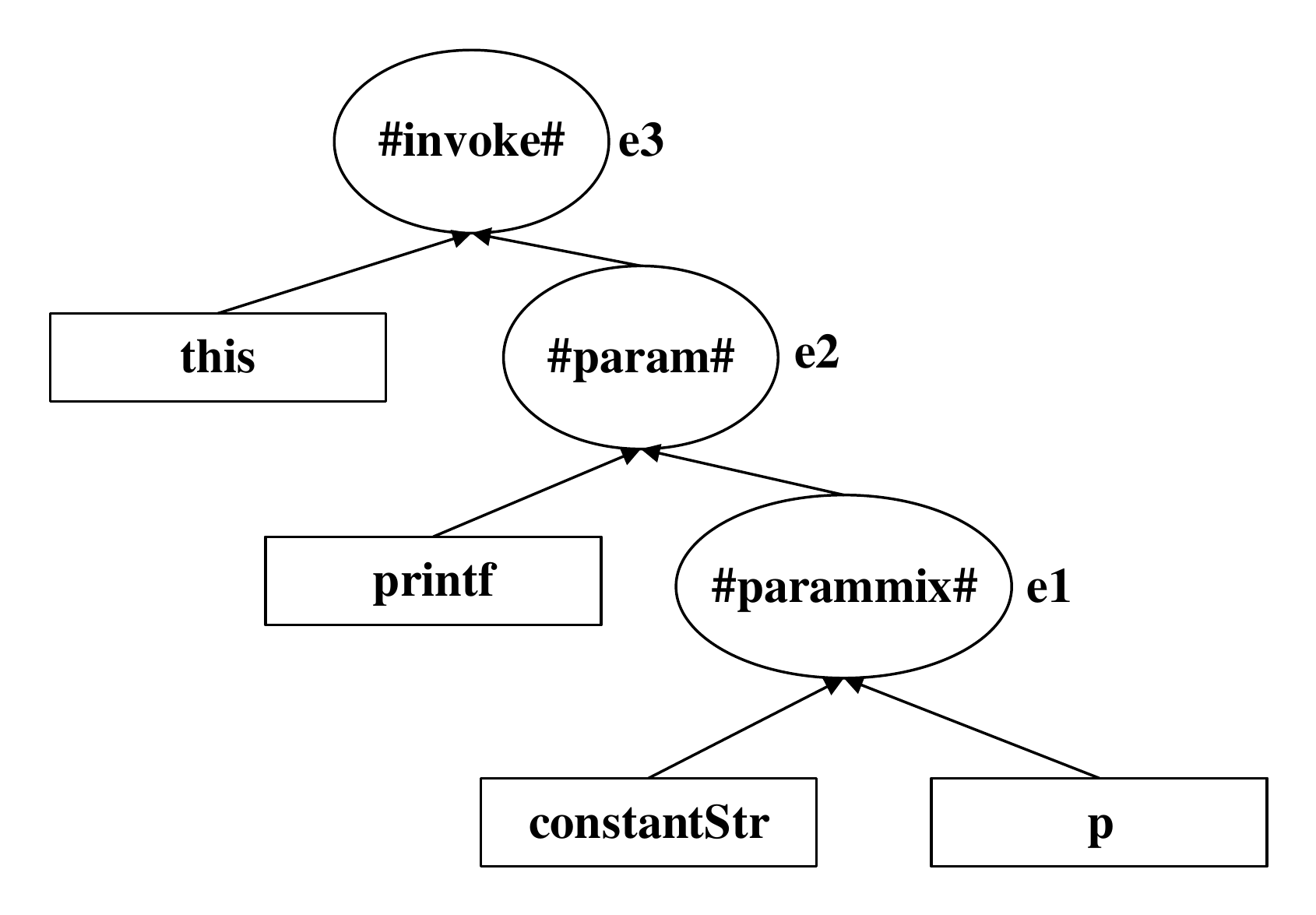}                    
	\caption{Event embedding tree}
	\label{fig:embedding_tree}        
	\vspace{-3mm}            
\end{figure}

In programming languages, a statement usually contains more than one event,
such as the statement this.printf("hello", p), which is composed of multiple
events. As shown in Fig. \ref{fig:embedding_tree}, the event embedding process
of the statement this.printf("hello", p) can be regarded as a tree structure.
The calculation process of event embedding needs to follow the rule of
calculating from the leaf nodes of the tree to the root node in turn. The
embedding process of the entire statement can be defined as follows:
\begin{equation}
\begin{split}
&e_1=embed(constantStr, \#parammix\#, p)\\
&e_2=embed(e_1, \#param\#, printf)\\
&e_3=embed(e_2, \#invoke\#, this)
\end{split}
\end{equation}
where $\#parammix\#$ represents the mixing of parameters before passing
parameters, $\#param\#$ represents passing parameters, $\#invoke\#$ represents
function calls, $e_1$, $e_2$ are intermediate events, and $e_3$ is the final
result of the event embedding of the statement. In this case, we say that event
$e_2$ depends on event $e_1$, and event $e_3$ depends on event $e_2$.

\begin{figure}[t]                                 
	\centering                                 
	\includegraphics[width=\linewidth]{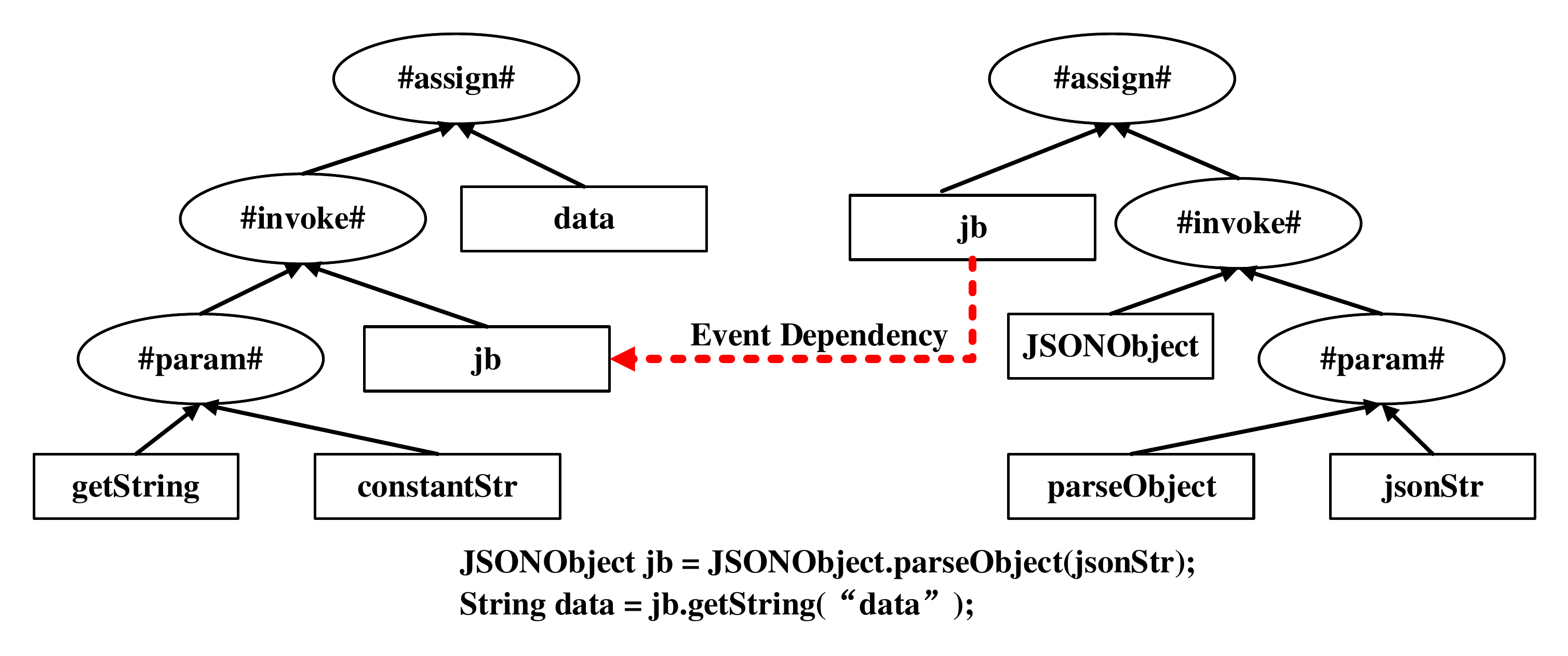}       
	\vspace{-7mm}             
	\caption{Event dependency in different statements}
	\label{fig:event_dependency}               
	\vspace{-6mm}         
\end{figure}
Note that the event dependency exists not only in a single program statement,
but also between multiple statements. As shown in Fig.
\ref{fig:event_dependency}, the embedding process of the statements shown in
the bottom of the figure can be transformed into two event embedding trees. In this
case, the event embedding of the second statement depends on the embedding
result of the first statement. Therefore, an event-dependent relationship
arises between the two statements. Our EDAM model can consider both these two
types of dependencies.  Note that the concept of event embedding tree is to
facilitate the explanation of the embedding calculation process of continuous
events. The dependency of events in the event embedding trees form an event
dependency graph, which is implemented in our EDAM model.

%implementation, we use the
%event dependency graph to describe the same calculation logic. The graphic
%structure is convenient for computer processing but not easy for people to
%understand. Therefore, we introduce the event embedding tree to explain the
%sequence of event embedding.

\section{model}\label{section:model}
\subsection{Overall Structure}

\begin{figure}[hbt]                                 
	\centering                                 
	\includegraphics[scale=0.5]{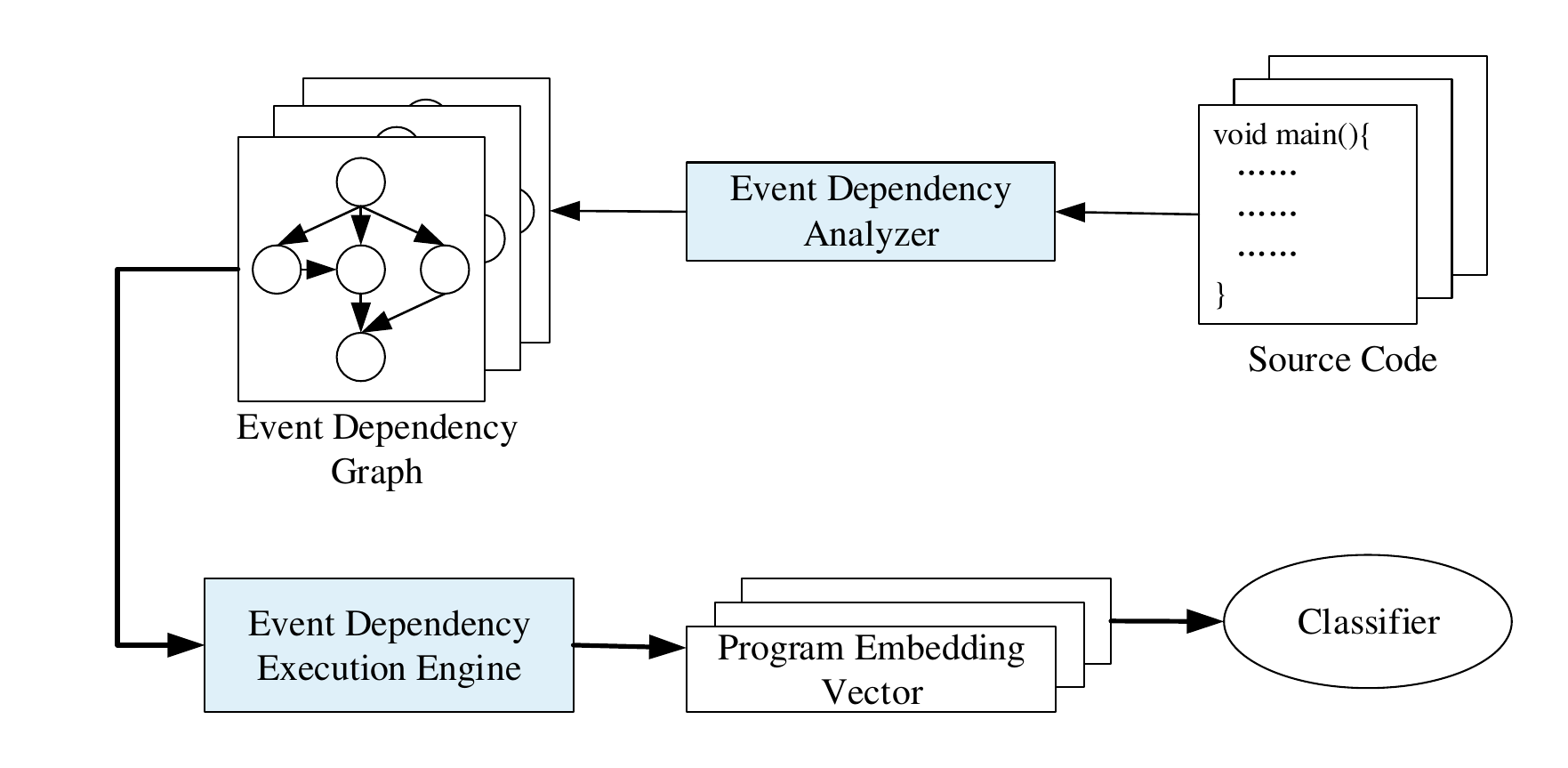}                    
	\caption{The structure of our model.}
	\label{fig:structure_model}     
	\vspace{-3mm}             
\end{figure}

As shown in Fig. \ref{fig:structure_model}, the core components of our model
are the event dependency analyzer and the event dependency execution engine,
where the event dependency analyzer is responsible for converting the source
code into an event dependency graph. The event execution engine is responsible
for converting the event dependency graph into a vector representation of the
program through calculation. The event dependency graph is a directed acyclic
graph. It describes not only the event dependency of each statement in the
program (as shown in Fig. \ref{fig:embedding_tree}), but also the event
dependency between different statements.  By analyzing the event information
provided in the event-dependent graph, the event-dependent execution engine
calculates the event semantics of the program to obtain the final program
semantic vector. The classifier is responsible for calculating the similarity
of the code pair according to the resulting program vector representation and
returning the classification result (i.e., whether it is a similar code pair).

\subsection{Preprocess}
In this section, we introduce the preprocessing process of the data set.

\noindent \textbf{Operator}:

For the C language, we identify 38 common operators (e.g, assign, return,
param, invoke, parammix, sizeof, structure access). These operators play the
role of P in the event triple $(A, P, O)$, and are used to describe the
relationship between two entities. For example, the operator in the event $c
\#<\# 1$ is $<$, which represents a numerical comparison between the variable
entity $c$ and the constant entity 1.

\noindent \textbf{Analysis of event dependence}

\begin{figure*}[ht]                                 
	\centering                                 
	\includegraphics[width=0.88\linewidth]{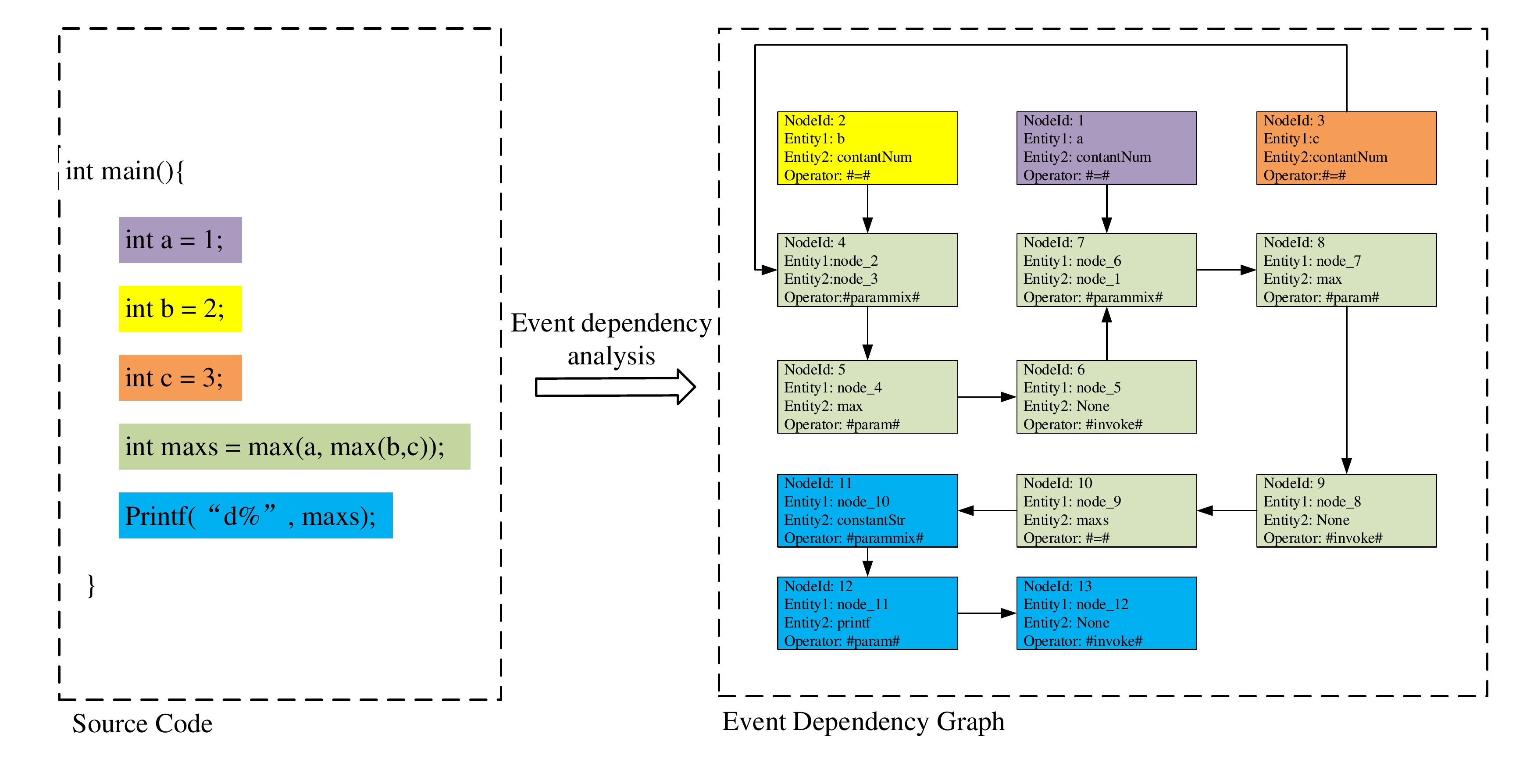}           
	\vspace{-4mm}         
	\caption{Event dependency graph}
	\label{fig:dependency_graph}         
	\vspace{-3mm}         
\end{figure*}

We use an event analyzer to analyze the event dependency in the program
according to the defined operator and generate an event dependency graph of the
program. There are two main types of event dependency in program code. The
first is the event dependency within a single program statement, and the second
is the event dependency between multiple program statements (e.g Section 3.2).
We use event dependency graphs to describe these two dependencies at the same
time.

As shown in Fig. \ref{fig:dependency_graph}, the left part of the figure is a
simple C language code fragment, and the right part is its corresponding event
dependency graph generated by the event dependency analyzer. In this code, the
event dependencies exist not only within a single statement, but also between
multiple statements. Each node in the event dependency graph represents an
event, and each edge from A to B represents that event B depends on event A.
According to the rules of our event dependency analyzer, each statement in the
code fragment on the left can be parsed into a subgraph in the code event
dependency graph on the right. The color of each statement in the left picture
corresponds to the color of the event-dependent subgraph of the corresponding
statement in the right picture. For example, the last statement printf("d\%",
maxs) in the left figure is marked in blue, which corresponds to the nodes 11,
12, and 13 that are also marked in blue in the event dependency graph.

The event dependency graph can not only represent the event dependency within a
single statement, but also reflect the event dependency between multiple
statements. For example, the printf statement uses the variable maxs, and maxs
is assigned in the previous statement. There is an event dependency
relationship between these two statements. This dependency relationship is
defined in the event dependency graph, that is, there is an edge from node 10
to node 11. The value of Entity1 in node 11 is node 10, which means that the
calculation result of node 10 will be represented as a vector of entity 1 in
node 11. In the event-dependent execution engine, we first use topological
sorting to split the event embedding tasks described by the event-dependent
graph to ensure that the event embedding nodes that need to rely on the
pre-node must be performed after the calculation of pre-node is completed.

\subsection{Event dependent execution engine}
\begin{figure*}[ht]                                 
	\centering                                 
	\includegraphics[width=0.86\linewidth]{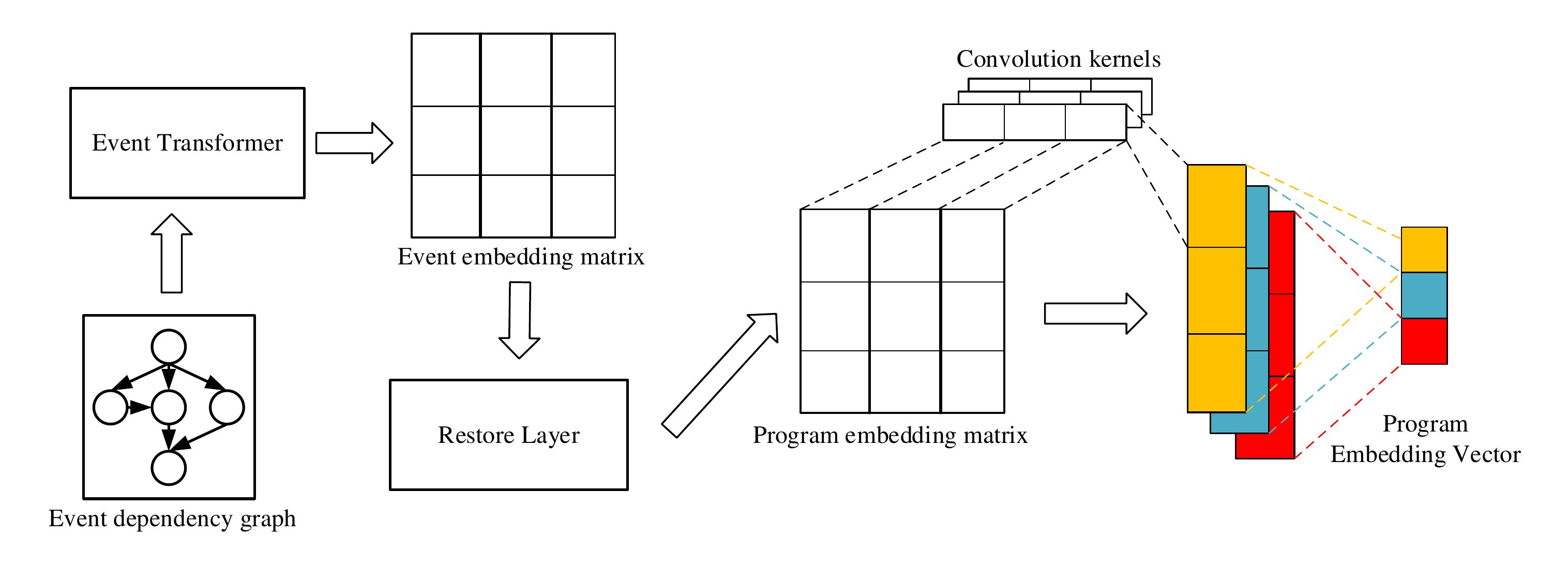}       
	\vspace{-3mm}             
	\caption{Execution process of event dependency execution engine.}
	\label{fig:execution_engine}       
	\vspace{-5mm}           
\end{figure*}

The event-dependent execution engine is responsible for performing calculations
based on the event-dependent graph to generate the semantic vector of the code.
The execution process of the event-dependent execution engine is shown in Fig.
\ref{fig:execution_engine}. The execution engine receives the event dependency
graph as input, and then uses the event Transformer defined in this paper to
embed the events described in the event dependency graph. The output of the
event transformer is a matrix composed of the event embedding vectors of each
node in the event dependency graph. We call this matrix the event embedding
matrix. The vector in the i-th row of the event embedding matrix is the event
embedding vector corresponded to node i in event dependency graph calculated by
Event Transformer. After that, we input the event embedding matrix into the
restore layer. The restore layer converts the event dependency matrix into a
program embedding matrix where the kth row of the program embedding matrix
represents the event embedding result of the kth row statement in the source
code. Finally, we use the convolutional layer to extract the semantics of the
program embedding matrix and generate the program embedding vector. In this
way, the program embedding vector contains the dynamic semantic information of
the source code fragment.

\begin{figure}[ht]                                 
	\centering                                 
	\includegraphics[width=\linewidth]{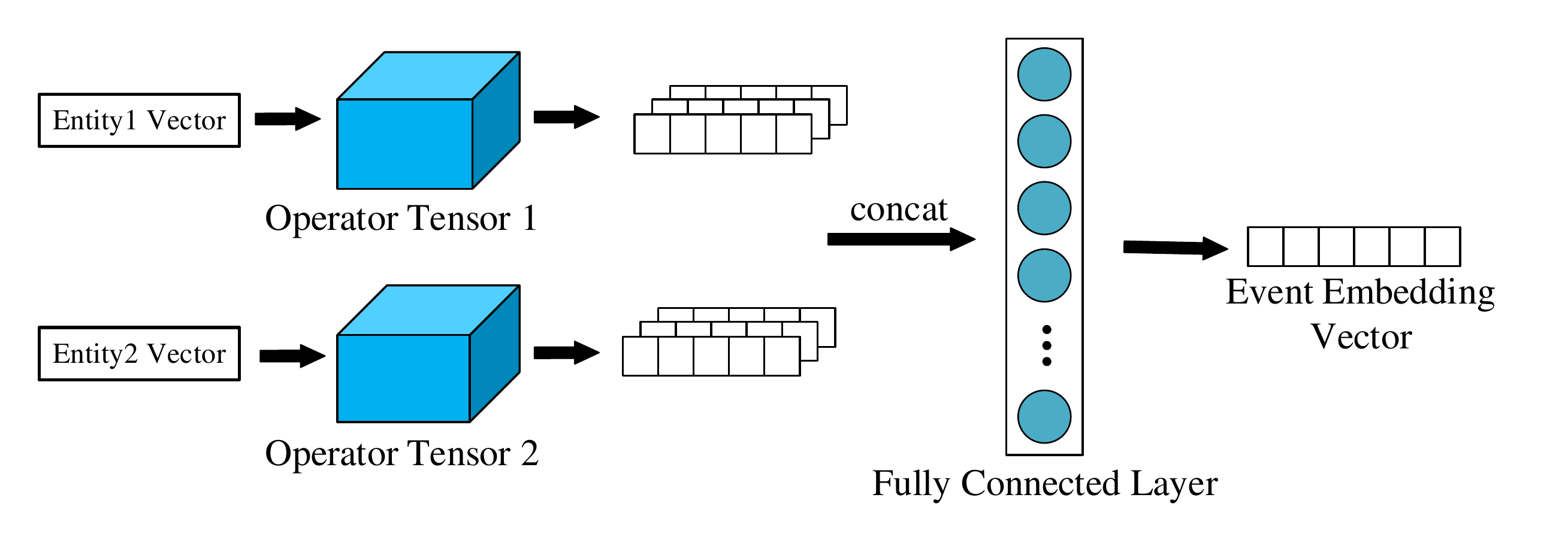}           
	\vspace{-4mm}         
	\caption{The structure of Event Cell}
	\label{fig:event_cell}
	\vspace{-7mm}          
\end{figure}

\noindent \textbf{Event Cell}

Event cell is the basic unit for event embedding calculation. Given a triple
$(A, P,O)$, the Event Cell is responsible for converting the event $(A, P, O)$
into the corresponding event embedding vector. The structure of Event Cell is
shown in Fig. \ref{fig:event_cell}. We define its calculation process as
follows: 
\begin{equation}
\begin{split}
e^k&=concat\left(vec\left(A\right)\ast T_{p1}^k,\ vec\left(O\right)\ast T_{p2}^k\right)\\
a&=concat(e^1,e^2,\cdots e^k)\\
o&=Dense(a)
\end{split}
\end{equation}
where $T_{p1}$ and $T_{p2}$ are two tensors with dimensions $(k, width,
height)$ associated with a specific operator $P$. These two tensors are
responsible for mapping entity vectors to multiple high-dimensional vector
spaces. $T_{p1}^k$ represents the k-th matrix in the tensor, and the size of
the matrix is $(width, height)$. $A$ and $O$ are entity 1 and entity 2
respectively. $Vec$ represents a function that maps entities $A$ and $O$ to
their corresponding vectors. The choice of the vec function can be diverse. One
can use models such as word2vec or BERT, or implement it in a custom way. The
$Dense$ function represents a fully connected layer.

In this paper, the $vec$ function we used treats different variable names and
function names as different entities. For example, printf (a, b, c) contains
four entities printf, a, b and c. Given a set of entities in a code fragment
$N={n_1,n_2,...,n_d}$, the vec function counts the number of times each entity
appears in the code fragment, and then the vec function renames the entity to
the form of $Top_i$ according to the number of times each entity appears in the
source code, where $i$ indicates that the number of occurrences of the entity
in the source code ranks i-th. Afterwards, the $vec$ function initializes
different $Top_i$ to a random entity vector.

In Event Cell, we use three-dimensional tensors $T_{p1}$, $T_{p2}$ instead of
two-dimensional matrices to map entity vectors to high-dimensional spaces. This
is mainly because three-dimensional tensors can map entity vectors to multiple
high-dimensional spaces, which can improve the semantic expression ability of
entity vectors.

\noindent \textbf{Event Transformer}

\begin{figure}[hbt]         
	\vspace{-5mm}                        
	\centering                                 
	\includegraphics[width=0.8\linewidth]{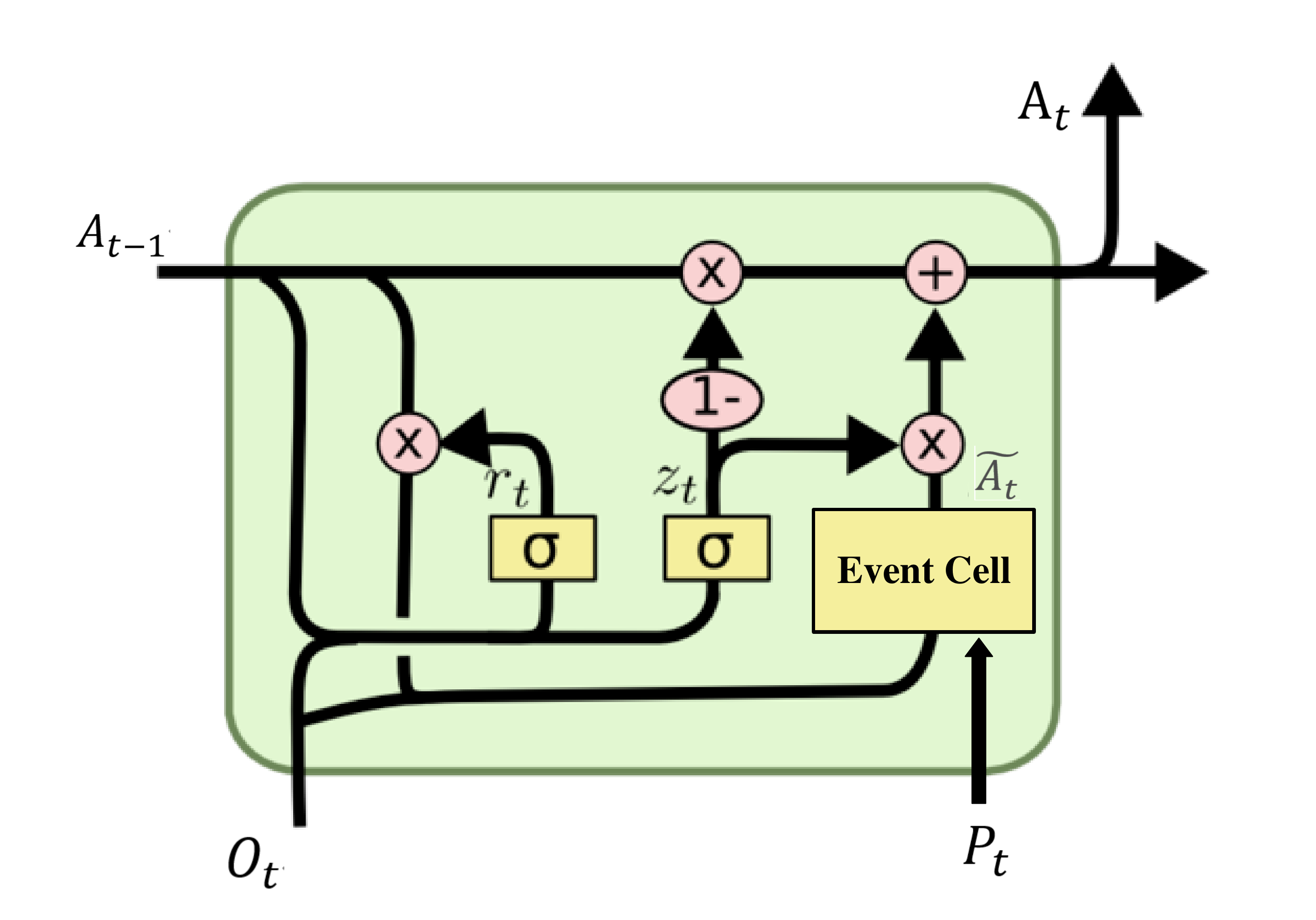}                    
	\caption{The structure of Event Transformer.}
	\label{fig:event_transformer}                  
\end{figure}

Event Cell can perform event embedding calculations on a single triple $(A, P,
O)$. The EventCell module can perform the event embedding calculation
of the program. However, this method has certain drawbacks. As shown in  Fig.
\ref{fig:dependency_graph}, the event embedding of node 13 depends on the event
embedding results of nodes 1 to 12, and these event embedding calculations
constitute an event embedding calculation chain. This chain structure causes us
to use Event Cell multiple times to recursively calculate each event on the
chain. This brings an obvious disadvantage: when the event embedding
calculation chain is very long, the early event information may be forgotten by
the model. Especially when a statement is very long and needs to rely on the
result of multiple pre-event embeddings, this problem becomes even more
important. Therefore, we introduce a gate mechanism to solve this
problem~\cite{li2020semantic}. The gate mechanism can enhance the model's
ability to remember early events. We introduce the gate mechanism into the
Event Cell and call it Event Transformer. The structure of Event Transformer is
shown in Fig. \ref{fig:event_transformer}.  We define the forward calculation
process of Event Transformer as follows:
\begin{equation}
\vspace{-1mm}
\begin{split}
r_t&=\sigma(W_r[A_{t-1},\ O_t])\\
z_t&=\sigma(W_z[A_{t-1},\ O_t])\\
\widetilde{A_t}&=Ec\left({r_t\ast A}_{t-1},P_t,O_t\right)\\
A_t&=\left(1-z_t\right)\ast A_{t-1}+z_t\ast\widetilde{A_t}
\end{split}
\end{equation}
where $W_r$ and $W_z$ are the weights of reset gate and update gate,
respectively. $A_{t-1}$ represents the calculation result of the previous time
step in the event embedding calculation chain. $E_c$ represents the Event Cell
modeule. $A_t$ is the calculation result of the current time step in the event
embedding calculation chain. $P_t$ is the operator corresponding to the current
time step.

\noindent \textbf{Restore Layer}
\begin{algorithm}[hbtp]
	\caption{Algorithm of restore layer}\label{algorithm}
	\label{arg:res}
	\KwData{event embedding matrix $E$ of size $w*l$}
	\KwResult{program embedding matrix $O$}
	$O \leftarrow []$;\\
	\For{$k \leftarrow 0$ \KwTo $w$}{
		currentNode $\leftarrow$ node withd id k in event dependency graph;\\
		\eIf{currentNode is the final event in a statement}{$O$.append($E[k]$);}{continue;}
	}
	return $O$;
\end{algorithm}

Through the above-defined Event Cell and Event Transformer, we can obtain the
event embedding matrix composed of the event embedding vector of each node in
the event dependency graph. In event embedding matrix, the i-th row vector of
matrix is the event embedding result of node with id i in event dependency
graph. Since multiple nodes of the event embedding graph usually only
correspond to one statement in the original code fragment, we need a restore
layer to convert the event embedding matrix into a program embedding matrix. In
the program embedding matrix output by the restoration layer, the kth row
vector of the matrix is the event embedding result of the kth statement in the
source code. The algorithm of the restore layer is defined as in
Algorithm.~\ref{arg:res}.

\noindent \textbf{Convolutional Layer}

The convolutional layer converts the program embedding matrix into a program
embedding vector, which is used for the subsequent code similarity calculation.
In the EDAM model, we use multiple convolution kernels and pooling layers to
perform this conversion. We define the calculation process of the convolution
layer as follows:
\begin{equation}
\begin{split}
\hat{X}&=Padding(X)\\
Z&=W_{conv}*\hat{X}^T\\
Q&=Pool(Z)
\end{split}
\end{equation}
where $X$ represents the program embedding matrix, we let $bz$ represent the
number of samples in each batch of training data, $n_o$ represents the number
of nodes in each program embedding matrix, $l_n$ represents the length of each
node vector of the event dependency graph, then the dimension of $X$ can be
expressed as $(bz, n_o, l_n)$. It should be noted that because the number of
nodes in different event dependency graphs is inconsistent, we need to padding
$X$ before proceeding to the next calculation. $\hat{X}$ represents the program
embedding matrix after padding. $W_{conv}$ is a matrix composed of multiple
one-dimensional convolution kernels, we let $n_k$ represent the number of
one-dimensional convolution kernels, $l_k$ represents the length of each
convolution kernel, and the shape of $W_{conv}$ can be expressed as $(n_k ,
l_k)$. Pool represents the average pooling operation on the last dimension of
$Z$. $Q$ is the program embedding vector output by the convolutional layer, and
its shape is $(bz, n_k)$. After passing through the convolutional layer, a
piece of source code can be expressed as an $n_k$ length program embedding
vector.

\subsection{Code Similarity Evaluation}
\vspace{-0.7mm}

Given the event embedding vectors of two pieces of code, we use cosine
similarity to evaluate the semantic similarity of the two pieces of code. In
the code clone detection system, we will calculate the similarity between the target
code and all candidate codes, and then determine whether the code pair is a
similar code that should be returned by the system according to the similarity
threshold $\theta$. When the cosine similarity of two pieces of code is greater
than $\theta$, we judge them as a similar code pair, and when the cosine
similarity is less than $\theta$, we judge them as a non-similar code pair.

\subsection{Parameter Learning}
We define the loss function as follows:
\begin{equation}
\begin{split}
g(x_i)=&Conv(Et(x_i))\\
sim(x_i,x_j)&=\frac{g(x_i) \cdot g(x_j)}{||x_i||*||x_j||}\\
Loss=\sum_{min}^{max} max(0&,1-sim(x_k,x_m)+sim(x_k,\hat{x}_k))
\end{split}
\end{equation}

Where $Et$ represents the Event Transformer module. $Conv$ represents the
convolutional layer, and $g(x_i)$ is the program embedding vector of the sample
$x_i$. $similairity$ represents a function for calculating the similarity of
two samples. In the Loss function, the code pair $(x_k, x_m)$ is a positive
sample, and $(x_k, \hat{x}_k)$ is a negative sample pair. We use random
sampling to collect negative samples. The experimental results show that the
use of negative samples in the model training process can effectively improve
the model's detection ability.

\subsection{Back Propagation Through Event(BPTE)}

BPTT algorithm is used in the traditional GRU unit training process because the
weights of reset gate, update gate and hidden layer can be shared through each
time step. In the calculation process of Event Transformer, $T_{p1}$ and
$T_{p2}$ in the Event Cell are determined by the operators in each embedded
step. Therefore, we propose the BPTE algorithm to train Event Transformer. The
back propagation process of BPTE algorithm is defined as follows:
\begin{equation}
\frac{\partial E_t}{\partial T_p}\ =\sum_{k\in N_p}\frac{\partial E_t}{\partial A_t}\ast\frac{\partial A_t}{\widetilde{A_t}}\left(\prod_{j=k+1}^{t}\ \frac{\partial\widetilde{A_j}}{\partial\widetilde{A_{j-1}}}\right)\ast\frac{\partial A_k}{\partial T_p}
\end{equation}
where $E_t$ is the training error at step t, and $N_p$ is the set of positions
where operator $p$ appears in the embedding tree. For example, for embedding $a
– b + c – d$, operator '$-$' appears in the embedding tree at positions 0 and
2, So $N_p = \{0, 2\}$. Through the BPTE algorithm, we can train the Event
Transformer. Since we have introduced a gate mechanism in the Event
Transformer, the Event Transformer can better fit the historical data in event
embedding calculation chain. Also, the exploding gradient and vanishing
gradient during the backpropagation can be addressed.

\section{Evaluation}\label{section:evaluation}
In this section, we evaluate our proposed model through experiments.\footnote{The dataset
	and the implementation of our model will be released in github.com after the paper is accepted.}. We mainly hope to answer the following questions through experiments:
\begin{itemize}
	\item \textbf{RQ1:} What are the advantages and disadvantages of our EDAM model compared to other commonly used models?
	
	\item \textbf{RQ2:} How does the similarity threshold affect the model prediction results?
	
	\item \textbf{RQ3:} Is it possible to improve the prediction accuracy of the model by simultaneously using multiple EDAM models for classification? Are these models complementary?
	
	\item \textbf{RQ4:} Can the program embedding vector express the semantic relevance of the code?
\end{itemize}

\subsection{Experiment Setup}
In this section, we introduce the details of setup in our expirement.

\noindent \textbf{Data Set}

In the experiment, we use the OJClone dataset. This open source dataset contains 104 OJ questions, each of which contains 500 user-submitted programming codes. Since each code fragment in the data set belongs to an independent problem, we regard the code fragments belonging to the same problem as functionally similar code pairs, and the code fragment belonging to different problems as non-similar code pairs. Also, the code fragments belonging the same question are submitted by different users, therefore they can be regarded as semantic similarity code clone pairs of Type-3/4. Without loss of generality, we select the 10 questions to evaluate our model, and for each question we randomly select 100 code fragments. In total, there are 1000 code fragments. We select 70\% of the samples in each problem to generate the training set, and the remaining 30\% of the samples to generate the test set. The training set contains 49,000 positive samples and 49,000 randomly sampled negative samples (96,000 in total). The test set contains 44850 samples, of which 4350 are positive samples and 40500 are negative samples.

\noindent \textbf{Metrics}

Three metrics are used to evaluate the performance of the models in the
experiments, i.e., Precision, Recall and F1 Score. The metrics are defined as follows:
\begin{equation}
\begin{split}
&Precision=\frac{TP}{TP+FP}\\
&Recall=\frac{TP}{TP+FN}\\
&F1=\frac{2*Precision*Recall}{Precision+Recall}
\end{split}
\end{equation}
%where $TP$ is the number of correctly classified as positive samples, $FP$ is
%the number of incorrectly classified as positive samples. $FN$ is the 
%number of positive samples misclassified as negative samples by the model.
%Precision mainly
%measures the accuracy of the model to classify positive samples. Recall mainly reflects the model's ability to cover the
%similar samples. The above two indicators are difficult to evaluate the true
%ability of the model when used alone, so in order to integrate the two
%evaluation indicators of Precision and Recall, the $F1$ Score is also introduced.
%This metric combines Precision and Recall, which can better evaluate the
%detection ability of the model.
where $TP$ represents the number of true positive samples, $FP$ represents the number of false positive samples. 
$FN$ represents the 
number of dalse negative samples.
Precision mainly
measures the accuracy of the model to classify positive samples. Recall mainly reflects the model's ability to cover the
similar samples. The above two indicators are difficult to evaluate the true
ability of the model when used alone, so in order to integrate the two
evaluation indicators of Precision and Recall, the $F1$ Score is also introduced.
This metric combines Precision and Recall, which can better evaluate the
detection ability of the model.

\subsection{Comparison of Different Models (RQ1)}
\begin{figure*}[hpbt]
	\centering                                 
	\includegraphics[width=\linewidth]{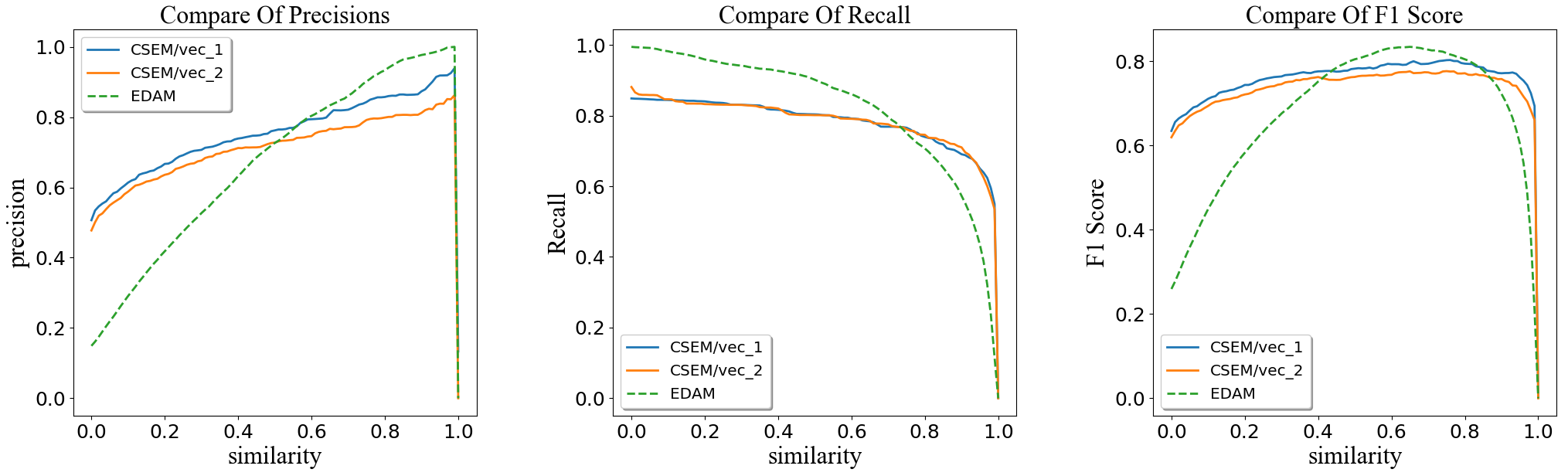}          
	\caption{Impact of Similarity Threshold.}
	\label{fig:similarity}
	\vspace{-5mm}                        
\end{figure*}
\begin{table}[hbtp]
	\vspace{-5mm}
	\caption{Comparison of Different Models.}
	\label{table:model_compare}
	\centering
	\renewcommand\arraystretch{2}
	\setlength{\tabcolsep}{1mm}{
		\begin{tabular}{|c|c|c|c|c|}
			\hline
			Model&Precision&Recall&F1 Score&Tool Configuration\\
			\hline
			CCLearner&0.0651&0.8324&0.1207&\makecell[c]{$\theta = 98\%,lr = 0.001,$\\ $epoch=300$}\\
			\hline
			Deckard&0.4201&0.3411&0.3765&\makecell[c]{$\theta = 95\%,MIT=30,$\\ $Stride=2$}\\
			\hline
			CloneWork&0.4604&0.3367&0.3890&\makecell[c]{$\theta=40\%,MIT=1,$\\$Mode=type3pattern$}\\
			\hline
			SourcerCC&0.5124&0.3210&0.3947&\makecell[c]{$\theta=50\%,$\\$MIT=1$}\\
			\hline
			CSEM/$vec_1$&0.8036&0.7192&0.7583&\makecell[c]{$\theta=80\%,K=2,$\\$tk=10, GA=[8,1]$}\\
			\hline
			CSEM/$vec_2$&0.7554&0.7293&0.7416&\makecell[c]{$\theta=70\%,K=2,$\\$GA=[8,1]$}\\
			\hline
			EDAM&\textbf{0.8029}&\textbf{0.8606}&\textbf{0.8307}&\makecell[c]{$\theta=70\%,K=2,$\\$GA=[8,1]$}\\
			\hline
	\end{tabular}}
\vspace{-3mm}
\end{table}
In this section, we compare the EDAM model with five state-of-the-art open source code clone detection models, which are CCLearner, Deckard, CloneWork, SourcerCC and CSEM. Among these models, Deckard, CloneWork, CSEM and SourcerCC can support the detection of C language code. The CCLearner model is a deep learning model for code feature extraction based on token. Its open source version only supports the detection of Java code. We have extended its code to support the processing of C language code. Since Type1/2 similarity codes are very easy to be detected, and almost all detection models can get good results, our experiment mainly focuses on the detection ability of Type3/4 code fragments.

The hyperparameter settings of these models are shown in Table.~\ref{table:model_compare}. When setting the hyperparameters of the comparison model and our EDAM model, we follow the principle of making the model's F1 Score perform best. In the table, $\theta$ represents the similarity threshold of the model, and lr represents the training learning rate. MIT is the shortest sequence length considered by the model. K is the length of the first dimension of the Operator Tensor in the Event Cell. GA indicates the number of multi-head attention in GAT layer (GA = [8, 1] indicates that there are two sub-layers in GAT layer, and the number of multi-head attention is setting to 8 and 1, respectively). For Deckard, we set its Stride to 2. For CloneWork, we set its detection mode to the highest type3pattern.

The experimental results are shown in Table. ~\ref{table:model_compare}. Our EDAM model is significantly better than other models in Precision, Recall, and F1 Score.We think this is because EDAM can better model the semantic information of the code. Compared with the CSEM model, which is also the code embedding model based on event embedding, our EDAM model still has a big improvement. We think this is mainly because we have introduced event-dependent information into the model. In our experiments, the performance of the CClearner model was poor. This is because the samples in our data set are homework codes submitted by students. The variable names in these code file  are relatively simple, and there are a large number of simple variable names and function names such as a, b, and c.The CCLearner model extracts code semantics based on tokens such as variable names and function names. These simple token names in the data set interfere with CCLearner's ability to correctly extract code semantic similarity, causing the CCLearner model to think that almost all code fragments are similar.

\subsection{Similarity Threshhold(RQ2)}
\begin{figure*}[hbpt]
	\centering
	\includegraphics[width=\linewidth]{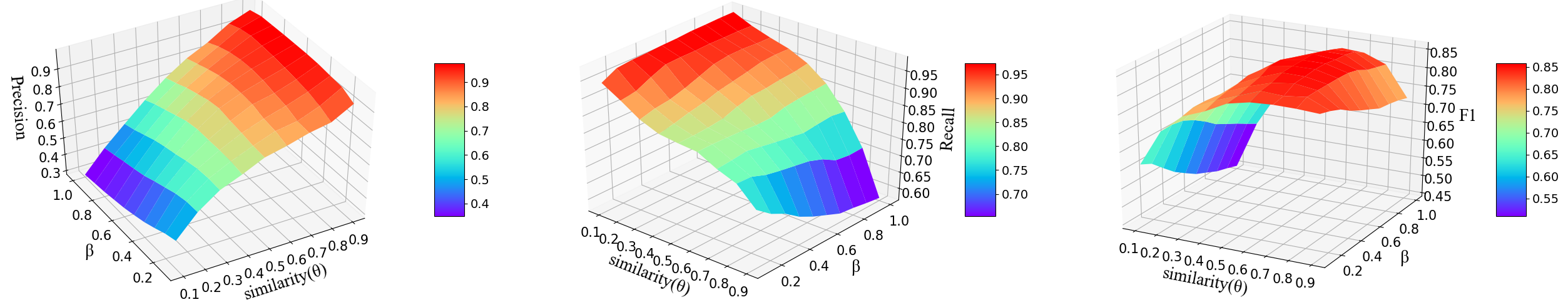}
	\caption{Impact of fusion parameters}
	\label{fig:fusion_parameter}
	\vspace{-5mm}
\end{figure*}
In this section, we mainly analyze the impact of different similarity thresholds on the model. We chose three models for comparison. The experimental results are shown in Fig.~\ref{fig:similarity}. The F1 Score of the EDAM model performs best when the similarity threshold is set to 60\%. The performance of the two sub-models of CSEM, CSEM/$vec_1$ and CSEM/$vec_2$ are relatively close, and CSEM/$vec_1$ is slightly better than CSEM/$vec_2$ in terms of precision and F1 Score. At the same time, we can observe an interesting phenomenon, that is, the slope of the first half of the curve of the CSEM model is relatively flat, and after reaching a certain threshold, the slope of the curve changes dramatically. This means that the program vector generated by the CSEM model is likely to appear in a small vector space, which is not convenient for the model to make further distinctions between positive and negative samples.The slope of the EDAM curve in the figure is obviously greater than the slope of the CSEM‘s, which indicates that the EDAM model is more sensitive to the similarity threshold than the CSEM model. We believe that this phenomenon reflects that the program embedding vectors generated by the EDAM model are distributed in a wider vector space. These vectors have higher cohesion when the categories are the same, and higher separation when the categories are different. In Section.~\ref{section:visualization}, we visualized the program embedding vectors generated by the EDAM model, and the results partially proved our above conclusions.

\subsection{Multi-model fusion(RQ3)}
\begin{table}[ht]
	\vspace{-3mm}
	\caption{Comparison of fusion models.}
	\label{table:model_fusion}
	\centering
	\renewcommand\arraystretch{2}
	\setlength{\tabcolsep}{1mm}{
	\begin{tabular}{|c|c|c|c|c|}
			\hline
			Model&Precision&Recall&F1 Score&Tool Configuration\\
			\hline
			CSEM/$f_1$&\textbf{0.9411}&0.6914&\textbf{0.7971}&$\theta_{vec_1} = 80\%, \theta_{vec_2} = 70\%$\\
			\hline
			CSEM/$f_2$&0.8856&0.7731&0.8255&$\theta=70\%,\beta=0.6$\\
			\hline
			EDAM/$f_1$&0.8818&\textbf{0.8486}&\textbf{0.8648}&$\theta=50\%,\beta=0.6$\\
			\hline
	\end{tabular}}
\vspace{-3mm}
\end{table}

\begin{figure}[htb]
	\centering
	\includegraphics[width=0.8\linewidth]{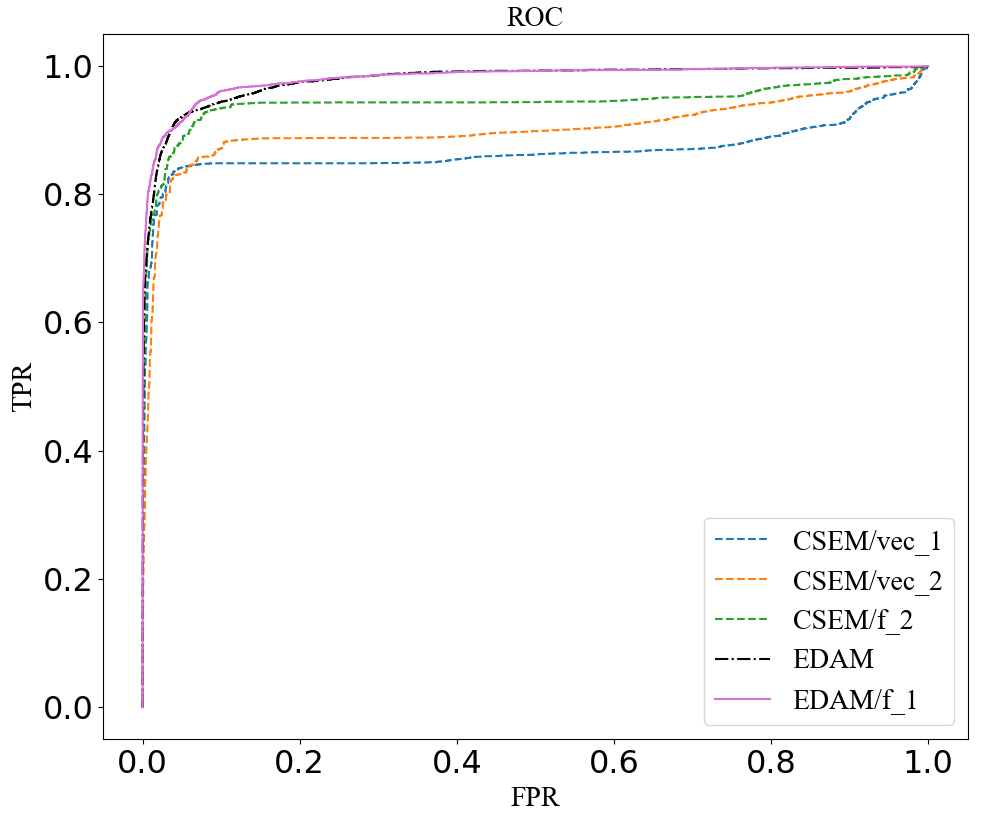}
	\vspace{-4mm}
	\caption{Roc of each model}
	\label{fig:roc}
	\vspace{-6mm}
\end{figure}

In the experimental process of this paper, we found that fusing the calculation results of multiple EDAM models can significantly improve the prediction accuracy of the model. This is because different EDAM models have different classification tendencies for the same samples due to random factors such as different sample input orders during the training process. Therefore, in this section, we propose an EDAM-based fusion model EDAM/$f_1$, and we analyze the difference in performance between the CSEM-based fusion models and the EDAM-based fusion models. The purpose of model fusion is to enhance the classification ability of the model by using the prediction results of multiple models at the same time.

The EDAM/$f_1$ model fuses the calculation results of two independent EDAM models, The formula is defined as $score(x1, x2) = \beta*score1(x_1, x_2) + (1-\beta)*score2(x_1, x_2)$, where $score1$ represents the score of the first EDAM model, $score2$ represents the score of the second EDAM model, and $\beta$ is used as a weight to balance the impact of the two models on the final score.

The experimental results are shown in Table.~\ref{table:model_fusion}, where CSEM/$f_1$ and CSEM/$f_2$ are two CSEM-based fusion models proposed by Li et al. The difference between the two lies in the way of fusion of the model. The CSEM/$f_1$ model performs best on Precisoion, while the EDAM/$f_1$ model performs best on Recall and F1 Score. This phenomenon indicates that the CSEM/$f_1$ model and EDAM/$f_1$ model have their own advantages in different types of tasks. When we need the similar samples predicted by the model to be as accurate as possible, the CSEM/$f_1$ model should be considered, because its Precision can reach 0.94, which means that when the CSEM/$f_1$ model judges a sample as a semantically similar sample, the reliability of the classification results is 94\%. Otherwise, when we need the model to detect as many similar samples as possible, we should consider using the EDAM/$f_1$ model, because its Recall reaches 0.84, which means that when the EDAM/$f_1$ model is used, 84\% of the similar samples in data set can be detected by the model.

We also compared the ROC of the models, and the experimental results are shown in Fig.~\ref{fig:roc}. The ROC of CSEM/$f_1$ is significantly better than CSEM/$vec_1$ and CSEM/$vec_2$ models.  After using multiple EDAM models for fusion, the EDAM/$f_1$ model has a further improvement compared to the EDAM model, so the EDAM-based fusion model EDAM/$f_1$ performs best among all models.

In order to further explore the influence of different fusion parameter settings on the EDAM/$f_1$ model, we analyzed the fusion parameter settings of the EDAM/$f_1$ model. The experimental results are shown in Fig.~\ref{fig:fusion_parameter}, where similarity($\theta$) and $\beta$  represent the similarity threshold and the balance parameter respectively. The z-axis of the three subgraphs represent the three evaluation metrics ,which are Precision, Recall, and F1. According to experiment result shown in Fig.~\ref{fig:fusion_parameter},  we can observe that when $\theta$ is set to 0.5 and $\beta$ is set to 0.6, the F1 Score of the model performs best.

\subsection{Visualization of code semantics(RQ4)}\label{section:visualization}
\begin{figure}[hbpt]
	\centering
	\includegraphics[width=0.6\linewidth]{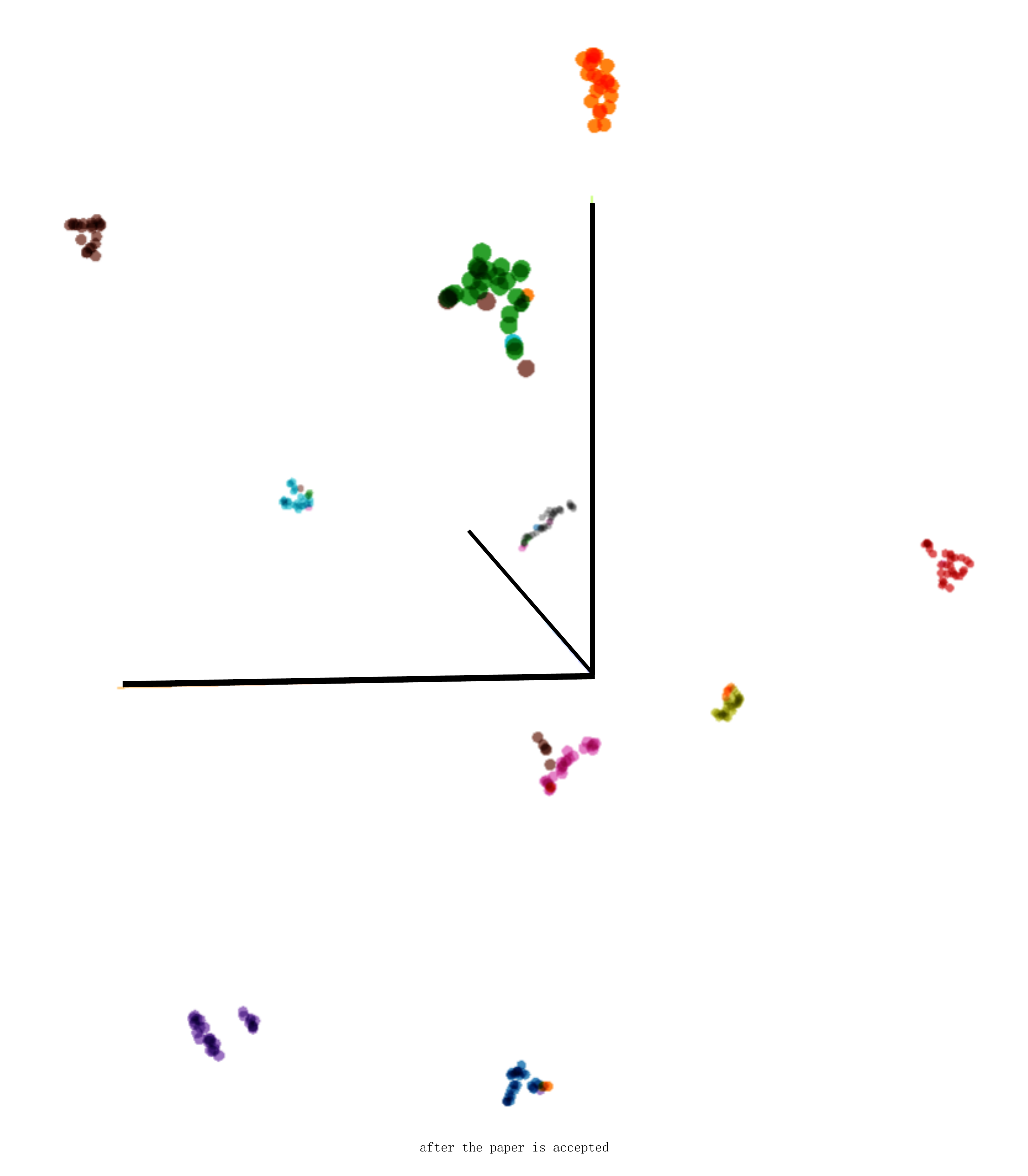}
	\caption{Visualization of program embedding vectors}
	\label{fig:visualization}
	\vspace{-2mm}
\end{figure}
In this section, we use the UMAP algorithm to visualize the program semantic embedding vector generated by the EDAM model. The purpose of this study is to verify whether the code semantic embedding vector generated by the model can show semantic relevance in the high-dimensional vector space. The experimental results are shown in Fig.~\ref{fig:visualization}. Each point in the figure represents a data point generated by mapping a program embedding vector to a three-dimensional space through the UMAP algorithm, and its color represents the sample category of the point. We used 10 categories of samples in the experiment, so the data points in the figure have 10 colors. We can observe that the data points in the graph are clustered into different clusters according to the different colors, while the data points of different colors are separated from each other. This result shows that our model can effectively learn the semantic information of the code, so that the code embedding vector generated by our model has good semantic expression ability in high-dimensional space.

\subsection{Threats to Validity}

The initialization of the network parameters and the negative samples selected during the training process will affect the performance of the model. To alleviate the influence of these confounding factors, we will train multiple models and use their average results to evaluate our model.

Uncertain factors in the model training process will also affect the performance of the model. In order to solve this problem and improve the performance of our model, we use a hybrid model to combine the advantages of the two models at the same time and improve the performance of our model.

In experiments, biased dataset may affect the generalization of the results. Therefore, in our experiment, we use the OJClone dataset to train our model. This dataset is an open source dataset containing many OJ questions, and each code fragment in the dataset is submitted by a different person. This means that the diversity of code semantics in the data set is rich, which is an important feature of real-world projects. Therefore, we choose this data set to solve the problem of bias in the data set.

\section{Conclusions and Future Work}\label{section:conclusion}

%\vspace{-1mm}
Detectiong Type-3/4 clone codes is important in many software engineering tasks and is crucial to the quality of software systems. In this paper, we propose an event embedding based method, EDAM, to detect semantic code clone.
We treat the program as a series of continuous interdependent events and propose a novel approach to model the execution semantics of statements by using event embedding and event dependency. In this way, both the execution semantics of
statements and their dependency infomation are captured to detect code fragments that are
similar in semantics but different in syntax. Experimental results show that our model outperforms
existing models in terms of Precision, Recall and F1 Score. In addition, we
demonstrate that the integration of two implementation options help to improve
the performance of our code clone detection model. We also conduct a visualization to
investigate the insights of our solution. 

Currently, our model supports C programs only. We plan to extend the model to
support Java and other languages. Also, we plan to conduct more experiments to
compare our model with other models using different dataset.

\bibliographystyle{IEEEtran}
\bibliography{reference.bib}

\end{document}